\documentclass[twocolumn,showpacs,showkeywords,prb]{revtex4}
\usepackage{graphicx}
\usepackage{dcolumn}
\usepackage{amsmath}
\usepackage{latexsym}

\def\JHEP{{\it JHEP} }

\def\NAT{{\it Nature} }

\def\PL{{\it Phys. Lett.} }
\def\PR{{\it Phys. Rev.} }

\begin{document}

\title{Time-Space Noncommutativity in Gravitational Quantum Well scenario}

\author{Anirban Saha}
\affiliation{Department of Physics, Presidency College,\\86/1 College Street, Kolkata-700073, India.}

\date{\today}

\begin{abstract}
\noindent
A novel approach to the analysis of the gravitational well problem from a second quantised description has been discussed. The second quantised formalism enables us to study the effect of time space noncommutativity in the gravitational well scenario which is hitherto unavailable in the literature. The corresponding first quantized theory reveals a leading order perturbation term of noncommutative origin. Latest experimental findings are used to estimate an upper bound on the time--space noncommutative parameter. Our results are found to be consistent with the order of magnitude estimations of other NC parameters reported earlier.
\end{abstract}

%{\bf{Keywords:}} Time-Space Noncommutativity, Gravitational Quantum Well.


\maketitle

\section{Introduction}
The idea of noncommutative (NC) space time where the coordinates $x^{\mu}$ satisfy the
noncommutative algebra
\begin{equation}
\left[x^{\mu}, x^{\nu}\right] = i \Theta^{\mu \nu}
\label{ncgometry}
\end{equation}
has gained prominence in the recent literature. Originally mooted by Snyder in a different perspective \cite{sny} this idea has been revived in the recent past \cite{sw} and field theories defined over this NC space are currently a subject of very intense research \cite{sz}. A wide range of theories are being formally studied in a NC perspective encompassing various gauge theories \cite{rbgauge} including gravity \cite{grav}.
Apart from studying the formal aspects of the NC geometry certain possible phenomenological consequences have also been investigated \cite{jabbari1, jabbari2, jabbari3, cs, rs1, rs2, rs3, rs4, rs5, rs6, rs7, rs8, rs9, rs10}.
A part of the endeavor is spent in finding the order of the NC parameter and in exploring its connection with observations \cite{cst, mpr, carol}.  

A particular piece of the scenario is the quantum well problem which has emerged in recent GRANIT experiments by Nesvizhevsky {\it et al.} \cite{nes1, nes2, nes3}  who detected the quantum states of the neutrons trapped in earth's gravitational field. Their results have been used by Bertolami {\it et al.} \cite{bert0, bert1} and Banerjee {\it et al.} \cite{RB} to set an upper bound on the momentum space NC parameters. These works have been done on the level of quantum mechanics (QM) where noncommutativity is introduced among the phase space variables. Naturally noncommutativity in the time-space sector can not be accounted for in this picture since in QM as such space and time could not be treated on an equal footing. Time-space noncommutativity, however, has gained considerable interest in current literature and a search for any possible upper bound on the time-space NC parameter using recent experimental feedback is very much desirable. 
 
 The issue of time-space noncommutativity is worth pursuing in  its own right because of its deep connection with such fundamental notions as unitarity and causality. It was argued that introduction of time-space noncommutativity spoils unitarity \cite{gomis, gaume} or even causality \cite{sei}. Much attention has been devoted in recent times to circumvent these difficulties in formulating theories with $\theta^{0i} \neq 0$ \cite{bala, bala1, Dayi, bert2}. In \cite{sch}, it was shown in the context of NC Schwinger model in $\left(1+1\right)$ - dimensions that in a perturbative approach retaining terms up to first order in the NC parameter does give physically meaningful results. There are similar examples of other theories with time-space noncommutativity in the literature \cite{luck, wood, ho1} where unitarity is preserved by an order by order perturbative approach. 

In the present paper we shall study the effect of time-space NC (if any) on the energy spectrum of a cold neutron trapped in a gravitational quantum well by restricting ourselves to first order perturbative treatment. To introduce time-space noncommutativity in this quantum well scenario a second quantized theory is required. We propose to discuss the NC quantum well problem reducing it from a NC Schr\"{o}dinger field theory. This is a reasonable starting point since single particle quantum mechanics can be viewed as the one-particle sector of quantum field theory in the very weakly coupled limit where the field equations are essentially obeyed by the Schr\"{o}dinger wave function \cite{Nair, RB1, bcsgas}. This allows us to examine the effect of the whole sector of space time noncommutativity in an effective noncommutative quantum mechanical (NCQM) theory. We do not consider momentum space NC effects as have been done by \cite{bert0, bert1, RB}. Our motivation is to study the effect of noncommutativity on the level of quantum mechanics when time-space noncommutativity is accounted for.

 The organization of the paper is the following. In the next section we consider a NC Schr\"{o}dinger field interacting with a external classical gravitational field. We show that the canonical structure of the effective commutative theory can be revived with suitable mass and field rescaling. Both the Lagrangean and Hamiltonian formulation is discussed. Once the canonical form is obtained we get back to the first quantized level in section 3. Here the ordinary quantum mechanics of the gravitational well problem is briefly reviewed before we consider the effective NCQM and work out the perturbed energy spectrum in three different approaches. Here, much to the spirit of \cite{bert0, RB}, we use the experimental results of \cite{nes1, nes2, nes3} to work out a estimation of the highest possible value of time-space NC parameter. In section 4 we make a rough calculation to show the consistency of our result with the estimations of other NC parameters existing in the literature \cite{bert0, RB}. We conclude in section 5.
 \vspace{-1.0 mm}
 %---------------------------------------------------------------------------
\section{The NC Schr\"{o}dinger action}
%---------------------------------------------------------------------------
In this section we shall consider a NC field theory of a nonrelativistic system with a constant background interaction. Evidently, the starting point is to write down the NC Schr\"{o}dinger action. Now there are two standard approaches to carry out the analysis of NC field theories. One can work in a certain Hilbert space which carries a representation of the basic NC algebra and the fields are defined in this Hilbert space by the Weyl--Wigner correspondence \cite{sz}.
We choose to take the alternative approach where we work in the deformed phase space with the ordinary product replaced by the star product \cite{sch, bcsgas, our}. In this formalism the fields are defined as functions of the phase space variables and the redefined product of two fields $\hat \phi(x)$ and $\hat \psi(x)$ is given by 
\begin{equation}
\hat \phi(x) \star \hat \psi(x) = \left(\hat \phi \star \hat \psi \right)(x) = e^{\frac{i}{2}
\theta^{\alpha\beta}\partial_{\alpha}\partial^{'}_{\beta}}
  \hat \phi (x) \hat \psi(x^{'})\big{|}_{x^{'}=x.}
\label{star}
\end{equation}
In star-product formalism the action for a NC Schr\"{o}dinger field $\hat \psi$ coupled with background gravitational field reads 
\begin{eqnarray} 
\hat S = \int dx \hspace{0.5mm} dy\hspace{0.5mm} dt \hspace{1.0mm} \hat \psi^{\dag}\star \left[i \hbar \partial_{0} + \frac{{\hbar}^{2}}{2m} \partial_{i}\partial_{i} - m g\hat{x} \right] \star \hat \psi
\label{NCaction} 
\end{eqnarray}
The above action describes a system in a vertical $xy$ ($i = 1, 2$) plane where the external gravitational field is taken parallel to the $x$-direction. 
Under $\star $ composition the Moyal bracket between the coordinates is
\begin{eqnarray} 
\left[\hat x^{\mu},\hat x^{\nu}\right]_{\star} = i\Theta^{\mu\nu} = \left(\begin{array}{ccc}
0 & -\eta & -\eta^{\prime} \\
\eta & 0 & \theta\\
\eta^{\prime}  & -\theta & 0\\
\end{array}\right)
\label{NCpara}
\end{eqnarray}
where $\mu, \nu$ take the values $0, 1, 2$. Spatial noncommutativity is denoted by $\Theta^{12} = \theta$ and noncommutativity among time and the two spatial directions are denoted by the parameters $\Theta^{10} = \eta$ and $\Theta^{20} = \eta^{\prime}$. 
 With the hindsight that any possible deformation in the ordinary theory due to noncommutativity is expected to be of small magnitude we expand the star product and consider only the first order correction terms, which considerably simplifies the analysis.
\vspace{-0.5cm}
%______________________________________________________
\subsection{First order equivalent commutative theory}
%______________________________________________________
Expanding the $\star$-product to first order in the NC parameters (\ref{NCpara}) we get 
\begin{eqnarray} 
\hat S = \int  dx \hspace{0.5mm} dy\hspace{0.5mm} dt \hspace{1.0mm} \psi^{\dag}\hspace{-4.0mm}  && \left[i \hbar \left( 1  - \frac{1}{2\hbar} m g \eta \right) \partial_{t}  +  \frac{{\hbar}^{2}}{2m} \partial_{i}{}^{2}\right. \nonumber\\
&&\left. - m g x  - \frac{i}{2} m g \theta \partial_{y} \right] \psi
\label{Caction} 
\end{eqnarray} 
where everything is in terms of commutative variables and NC effect is manifest 
by the presence of $\theta$ and $\eta$ terms.
Clearly the standard form of the kinetic term of Schr\"{o}dinger action is deformed due to time-space noncommutativity. To make the matters simpler we rescale the field variable by 
\begin{eqnarray}
\psi \mapsto \tilde{\psi} = \sqrt{\left(1 - \frac{\eta}{2 \hbar} m g\right)}\hspace{2.0 mm} \psi
\label{scal1}
\end{eqnarray}
which gives conventionally normalized kinetic term. 
Such physically irrelevant rescalings have been done earlier \cite{bcsgas, carol}.
Therefore it becomes clear that it is $\tilde \psi$, rather than $ \psi$, which corresponds to the  basic field variable in the action (\ref{Caction}). It is therefore desirable to re-express it in terms of $\tilde\psi$ and ensure that it is in the standard form in the first pair of terms. 
\begin{eqnarray} 
\hat S = \int  dx \hspace{0.5mm} dy\hspace{0.5mm} dt \hspace{1.0mm} \tilde{\psi}^{\dag}\hspace{-4.0mm}  && \left[i \hbar \partial_{t}  +  \frac{{\hbar}^{2}}{2\tilde{m}} \partial_{i}{}^{2} - \tilde {m}\left( 1  +  \frac{\tilde{m} g \eta}{\hbar} \right)  g x \right. \nonumber\\
&&\left.  - \frac{i}{2} \tilde{m} g \theta \partial_{y} \right] \tilde{\psi}
\label{Caction1} 
\end{eqnarray} 
Note that the mass term has also been rescaled 
\begin{eqnarray}
\tilde m = \left(1 - \frac{\eta}{2 \hbar} m g \right) m.
\label{scale2}
\end{eqnarray}
and we can interpret $\tilde {m}$ as the observable mass. A similar charge rescaling of NC origin in context of NC QED was shown in \cite{carol}. The last term in(\ref{Caction1}) can be absorbed in the $\partial_{y}{}^{2}$ by rewriting
\begin{eqnarray}
\partial_{y} = \left(\partial_{y} -  \frac{i \theta}{2 \hbar^{2}} \tilde {m}{}^{2} g \right) .
\label{scale3}
\end{eqnarray}
and the final effective NC Schr\"{o}dinger action reads 
\begin{eqnarray} 
\hat S = \int  dx \hspace{0.5mm} dy\hspace{0.5mm} dt \hspace{1.0mm} \tilde{\psi}^{\dag}\hspace{-4.0mm}  && \left[i \hbar \partial_{t}  +  \frac{{\hbar}^{2}}{2\tilde{m}} \left(\partial_{x}{}^{2} + \partial_{y}{}^{2}\right)\right.\nonumber\\
&&- \left.\tilde {m}g x  -  \eta \left(\frac{\tilde{m}{}^{2} g^{2}}{\hbar}\right)x \right] \tilde{\psi}
\label{Caction2} 
\end{eqnarray} 
The Lagrange equation of motion for the fundamental field $\tilde {\psi}(x)$ is
\begin{eqnarray} 
\left[i \hbar \partial_{t}  +  \frac{{\hbar}^{2}}{2\tilde{m}} \left(\partial_{x}{}^{2} + \partial_{y}{}^{2}\right) - \tilde {m}g x  -  \eta \left(\frac{\tilde{m}{}^{2} g^{2}}{\hbar}\right) x \right] \tilde{\psi} = 0 \nonumber\\
\label{eqm} 
\end{eqnarray} 
Note that owing to the field and mass redefinition (\ref{scal1}, \ref{scale2}) everything but the last term takes the form of standard Schr\"{o}dinger field equation.  
%________________________________
\subsection{Hamiltonian Analysis}
%________________________________
We can also start with the commutative equivalent action (\ref{Caction}) and derive the field equation in a Hamiltonian formalism. The advantage of the Hamiltonian analysis is twofold. It gives us a suitable platform to look for the proper canonical pair of fields as well as a crosscheck for the field equations.

The canonical momenta corresponding to the field variable $\psi$ is 
\begin{eqnarray}
\Pi_{\psi} \left(x\right) =  i \hbar \left( 1  - \frac{1}{2\hbar} m g \eta \right) \psi^{\dag}\left(x\right)
\label{momenta}
\end{eqnarray}
Note that in the argument we collectively refer to both the spatial coordinates by $x$. 
Writing the usual Poission bracket (PB) between the canonical pair
\begin{eqnarray}
\left\{\psi(x), \Pi^{\dag}(x^{\prime})\right\} = \delta^{2}(x - y)
\label{cononical}
\end{eqnarray}
and using the Faddiv--Jackiw technique \cite{FJ} we get the basic bracket, given by 
\begin{eqnarray}
\left\{ \psi(x), \psi^{\dag}\left( x^{\prime}\right)\right\} = - \frac{i}{\hbar}\left( 1  + \frac{1}{2\hbar} m g \eta \right)  \delta^{2}\left(x -  x^{\prime}\right)
\label{commutation}
\end{eqnarray}
This is the non-standard form of the PB relation and as such indicates that $\psi$ cannot represent the basic field variable. Following a physically trivial field rescaling (\ref{scal1}) we get the usual PB structure:
\begin{eqnarray}
\left\{\tilde \psi(x), \tilde \psi^{\dag}\left( x^{\prime}\right)\right\} = - \frac{i}{\hbar} \delta^{2}\left(x -  x^{\prime}\right)
\label{commutation}
\end{eqnarray}
which justifies our earlier argument that instead of the original fields $\psi$ one should choose the rescaled field variables $\tilde{\psi}$ and $\tilde{\psi}^{\dag}$ as the canonical pair of fields. 

 The Hamiltonian density is worked out using (\ref{Caction}) and (\ref{momenta})
\begin{eqnarray}
{\cal{H}} & = & \Pi_{\psi} \dot{\psi} - {\cal{L}}  \nonumber\\
& = & - \frac{{\hbar}^{2}}{2m} \psi^{\dag}\partial_{i}{}^{2}\psi + m g \psi^{\dag} x \psi   + \frac{i}{2} m g \theta \psi^{\dag}\partial_{y}\psi
\label{Ham}
\end{eqnarray}
and rewritten in terms of the rescaled fields $\tilde{\psi}, \tilde{\psi}^{\dag}$ and mass $\tilde{m}$:
\begin{eqnarray} 
{\cal{H}} \left(x\right)=  - \frac{{\hbar}^{2}}{2\tilde{m}} \tilde{\psi}^{\dag}\partial_{i}{}^{2}\tilde{\psi}  + \tilde{m} g \left(1 + \eta \frac{\tilde{m} g}{\hbar}\right)\tilde{\psi}^{\dag} x \tilde{\psi}
\label{Ham1} 
\end{eqnarray} 
where the last term in (\ref{Ham}) is absorbed in $\partial_{y}{}^{2}$ as usual (\ref{scale3}).
This Hamiltonian density generates the time evolution of the system 
\begin{eqnarray} 
\dot{\tilde{\psi}} & = & \left\{\tilde{\psi}\left(x\right), {\cal{H}} \left(x^{\prime}\right)\right\}\nonumber\\
& = &  -\frac{i}{\hbar}\left[- \frac{{\hbar}^{2}}{2\tilde{m}} \partial_{i}{}^{2}  + \tilde{m} g \left(1 + \eta \frac{\tilde{m} g}{\hbar}\right) x\right] \tilde{\psi}
\label{Ham_eqn} 
\end{eqnarray} 
which is same as our Lagrange field equation (\ref{eqm}). \\
%%%%%%%%%%%%%%%%%%%%%%%%%%%%%%%%%%%%%%%%%%%%%%%%%%%%%%%%%%%%%%%%%%%%%%%%%%%%%%%%
\section{Reduction to first quantized theory}
%%%%%%%%%%%%%%%%%%%%%%%%%%%%%%%%%%%%%%%%%%%%%%%%%%%%%%%%%%%%%%%%%%%%%%%%%%%%%%%%
So far we have been dealing with the second quantized formalism where $\tilde{\psi}$ was the basic field variable of the theory. The motivation was to impose non commutativity in the time-space sector and study how it affects the system. We found out that the only nontrivial change in the Schr\"{o}dinger equation is indeed originating from the space--time noncommutativity. Specifically, it shows up only in the direction of the external gravitational field ${\bf{g}} = -\rm{ g {\bf {e_{x}}}}$. This result is in conformity with \cite{bert0, bert1, RB} where it is shown that it is momentum noncommutativity, not space coordinate noncommutativity that shows up in first order computations. In \cite{bert0, bert1, RB} along with spatial noncommutativity, momentum space noncommutativity has been included as well. However their treatment essentially leaves a gap in the analysis which we fill in here. 
Since first and second quantized formalisms are equivalent as far as Galilean systems are concerned, in the sequel of the paper we carry out a equivalent NC quantum mechanical analysis in the first quantized formalism. 

% We now turn our attention to
 In the first quantized version of the theory the Schr\"{o}dinger field equation (\ref{eqm} or \ref{Ham_eqn}) will be treated as the quantum mechanical equation of motion and the earlier field variable $\tilde{\psi}$ will be interpreted as wave function. This is a quick and simple but standard procedure to reduce the field theoretic setup to one-particle quantum mechanics as has been illustrated in \cite{Nair} for a general external potential.

 We begin by checking that $\tilde {\psi}$ does have an interpretation of probability amplitude and satisfies the continuity equation 
\begin{eqnarray}
\partial_{0} j_{0} + \partial_{i} j_{i} = 0 ; \qquad (i=1,2) 
\label{continuity}
\end{eqnarray}
with the usual expressions of probability density $j_{0} $ and probability current $j_{i}$ in terms of $\tilde {\psi}$.
From equation (\ref{Ham_eqn}) we easily read off the Hamiltonian as
\begin{eqnarray} 
H = H_{0} + H_{1} = \frac{1}{2\tilde{m}} \left(p_{x}{}^{2} + p_{y}{}^{2}\right) + \tilde {m}g x  +  \eta \frac{\tilde{m}{}^{2} g^{2}}{\hbar} x \nonumber\\
\label{H} 
\end{eqnarray} 
Note that the the NC effect in the ordinary part $H_{0}$ is hidden in the mass and field redefinition (\ref{scal1}, \ref{scale2}). Such rescalings are only viable in a region of space time where variation of the external field is negligible. Since the results we have derived are to be compared with the outcome of a laboratory-based experiment we can safely assume a constant external gravitational field throughout.

Before proceeding with the Hamiltonian (\ref{H}) we should note that even if the variables in the commutative equivalent model are commuting it is not obvious that the usual Hamiltonian procedure could produce dynamics with respect to noncommuting time. To first order in the NC parameter it has been shown in \cite{Dayi} that time space noncommutativity emerges from a duality transformation and Hamiltonian analysis is identical for both the original theory ( with noncommutativity in the spatial sector only ) and its dual containing space time noncommutativity. Such approach has been shown to lead to reasonable outcome in \cite{sch}. Following these we propose to carry out our analysis to first order in $\eta$ and assume the applicability of the usual Hamiltonian dynamics for the commutative equivalent model. 

  Since we expect the time-space NC parameter is rather small at the quantum mechanical level, the last term in equation-(\ref{H}) represents a perturbation $H_{1}$ in the usual gravitational quantum well scenario described by $H_{0}$.  We now briefly review the ordinary quantum well problem, its solutions and the experimental results \cite{nes1, nes2} before any further discussion of the NC extension. 
%____________________________________________________
\subsection{Ordinary Gravitational Quantum Well}
%____________________________________________________
The first two terms in (\ref{H}) is the commutative Hamiltonian $H_{0}$ of the gravitational well problem which describes the quantum states of a particle with mass $\tilde{m}$ trapped in a linear potential well, in this case a gravitational well.
The system's wave function can be separated into two parts, corresponding to each of the coordinates $x$ and $y$. Since the particle is free to move in $y$-direction its energy spectrum is continuous along $y$ and the corresponding wave function can be written as collection of plane waves 
\begin{eqnarray} \tilde{\psi}(y)=\int_{-\infty}^{+\infty}g(k)e^{i k y}dk~,
\label{Airy_5}
\end{eqnarray}
where the function $g(k)$ determines the group's shape in phase space.
The analytical solution of the Schr\"{o}dinger equation in $x$ direction 
%We consider now the experiment described in Refs. \cite{Nesvizhevsky_1, Nesvizhevsky_2}. In the case where a horizontal mirror is placed at $x=0$, a quantum well is formed by the mirror and the constant gravitational field. This system is known as the \emph{gravitational quantum well}.
i.e. the eigenvalue equation , $H_{0}\tilde{\psi}_n=E_n\tilde{\psi}_n$, are well known \cite{Landau}. The eigenfunctions corresponding to $x$ can be expressed in terms of the Airy function $\phi(z)$,
\begin{eqnarray}
\psi_n(x)=A_n\phi(z)~,
\label{Airy_1}
\end{eqnarray}
with eigenvalues determined by the roots of the Airy function, $\alpha_n$, with $n=1,2\ldots$,
\begin{eqnarray}E_n=-\bigg({\tilde{m}g^2\hbar^2\over2}\bigg)^{1/3}\alpha_n~.
\label{Airy_2}
\end{eqnarray}
The dimensionless variable $z$ is related to the height $x$ by means of the following linear relation:
\begin{eqnarray}z=\bigg({2m^2g\over\hbar^2}\bigg)^{1/3}\bigg(x - {E_n\over\tilde{m}g}\bigg)~.
\label{Airy_3}
\end{eqnarray}
The normalization factor for the $n$-th eigenstate is given by:
\begin{eqnarray} A_n=\Bigg[\bigg({\hbar^2\over 2m^2g}\bigg)^{1\over3}\int_{\alpha_n}^{+\infty}dz\phi^2(z)\Bigg]^{-{1\over2}}~.
\label{Airy_4}
\end{eqnarray}
The wave function for a particle with energy $E_{n}$ oscillates below the classically allowed hight $x_{n} = \frac{E_{n}}{\tilde{m}g}$ and above $x_{n}$ it decays exponentially. This was realized experimentally by Nesvizhevsky {\it{et al}} \cite{nes1, nes2, nes3} where they observed the lowest quantum state of neutrons in the earth's gravitational field. The idea of the experiment was to let cold neutrons flow with a certain horizontal velocity ($6.5\:\mathrm{ms^{-1}}$) through a horizontal slit formed between a mirror below and an absorber above. The number of transmitted neutrons as a function of absorber hight is recorded and the classical dependence is observed to change into a stepwise quantum-mechanical dependence at a small absorber hight.
Their results and a comparison with the theoretical values are given below.
%Recently, Nesvizhevsky \emph{et al.} \cite{Nesvizhevsky_2} were able to determine the
The experimentally found value of the classical height for the first quantum state is:
\begin{eqnarray} 
x_1^{exp}&=&12.2\pm1.8(\mathrm{syst.})\pm0.7(\mathrm{stat.})\ (\mathrm{\mu m})
%~,\nonumber\\ x_2^{exp}&=&21.6\pm2.2(\mathrm{syst.})\pm0.7(\mathrm{stat.})\ (\mathrm{\mu m})~.
\label{Nesvizhevsky_1}
\end{eqnarray}
The corresponding theoretical value can be determined from Eq. (\ref{Airy_2}) for $\alpha_1=-2.338$ 
% and $\alpha_2=-4.088$, 
yielding 
\begin{eqnarray} 
x_1=13.7\:\mathrm{\mu m}
%~,\mathrm{for}~ E_1=1.407\:\mathrm{peV}
%\nonumber\\ x_2=24.0\:\mathrm{\mu m}~,\mathrm{for}~ E_2=2.461\:\mathrm{peV}.
\label{Nesvizhevsky_1a}
\end{eqnarray}
This value is contained in the error bars and allow for maximum absolute shift of the first energy level with respect to the predicted values:
\begin{eqnarray} 
\Delta E_1^{exp}&=&6.55\times10^{-32}\ \mathrm{J}=0.41\ \mathrm{peV}
%\nonumber\\ \Delta E_2^{exp}&=&8.68\times10^{-32}\ \mathrm{J}=0.54\ \mathrm{peV}~.
\label{Nesvizhesky_2}
\end{eqnarray}
The values of the constants taken in this calculations are as follows:
\begin{eqnarray} 
\hbar & = & 10.59\times 10^{-35}~\mathrm{Js}\nonumber\\
g & = & 9.81~ \mathrm{ms^{-2}}\nonumber\\
\tilde{m} & = & 167.32 \times 10^{-29}~\mathrm{Kg}.
\label{constants}
\end{eqnarray}

%-----------------------------------------------------
\subsection{Analysis of the perturbed energy spectrum}
%-----------------------------------------------------
Going back to the effective NCQM theory we now analyze the perturbed system \ref{H}. The perturbative potential is given by 
\begin{eqnarray}
H_{1} = \eta\left(\frac{{\tilde{m}}^{2}{g}^{2}}{\hbar}  \right)x
\label{perturbation}
\end{eqnarray}
Interestingly the occurrence of this perturbation term is a direct manifestation of time--space noncommutativity. This enables us to work out a upper bound for the time--space noncommutative parameter. Following the prescription of \cite{bert0} we can demand that the correction due to (\ref{H}) in the energy spectrum should be smaller or equal to the maximum energy shift allowed by the experiment \cite{nes1, nes2, nes3}. We work out the theoretical value of the energy shift in three independent ways.
%_______________________________
\subsubsection{Numerical Method}
%_______________________________
First we take the numerical approach similar to \cite{bert0} and calculate the leading order energy shift of the first quantum state. It is just the expectation value of the perturbation potential, given by 
\begin{eqnarray}
\Delta E_{1}&=&\eta \frac{\tilde{m}^{2}g^{2}}{\hbar}
\int_0^{+\infty} dx~\tilde{\psi}_{1}^{*}(x)~x~\tilde{\psi}_{1}(x)
\nonumber\\
&=& \eta \frac{\tilde{m}^{2} g^{2}}{\hbar} \Bigg[\bigg({2\tilde{m}{}^2g\over\hbar^2}\bigg)^{-\frac{2}{3}}A_{1}^2I_{1}+ {E_{1}\over \tilde{m}g}\Bigg]~,
\label{energy}
\end{eqnarray}
where the integral $I_{1}$ is defined as:
\begin{eqnarray} 
I_{1}\equiv\int_{\alpha_{1}}^{+\infty}dz\phi(z)z\phi(z)~,
\label{int}
\end{eqnarray}
The values of the first unperturbed energy level $E_{1}$ is determined from (\ref{Airy_2}) with $\alpha_{1} = -2.338 $ :
\begin{eqnarray} 
E_{1}  =  2.259 \times 10^{-31} \ \mathrm{(J)}  = 1.407 \ \mathrm{(peV)}
\label{energy_1}
\end{eqnarray}
The normalization factor $A_{1}$ is calculated from (\ref{Airy_4}). The integrals in  (\ref{Airy_4}) and (\ref{int}) were numerically determined for the first energy level:
\begin{eqnarray} 
A_{1}  =  588.109\ ,\qquad I_{1}  =  -0.383213~
%&A_2= 513.489\ ,\qquad & I_2=-0.878893~.
\label{numerical_1}
\end{eqnarray}
The first order correction in the energy level is given by 
\begin{eqnarray} 
\Delta E_{1} = 2.316\times10^{-23} \eta\quad\mathrm{(J)~,}
%\Delta E_2^{(1)}=4.94\times10^{29}\eta\qquad\mathrm{(J)~.}
\label{energy_corrections}
\end{eqnarray}
Comparing with the experimentally determined value of the energy level from (\ref{Nesvizhesky_2}) we found the bound on the time--space NC parameter is 
\begin{eqnarray} 
|\eta| & \lesssim & 2.83\times 10^{-9}\ \mathrm{m^{2}}
%|\eta|&\lesssim& 1.76\times 10^{-61}\ \mathrm{kg^2m^2s^{-2}}\qquad(n=2)~.
\label{eta_bounds_1}
\end{eqnarray}
%_________________________
\subsubsection{WKB Method}
%_________________________
Avoiding the numerical methods one can analyze the energy spectrum using a quasiclassical approximation. The potential term in the unperturbed Hamiltonian $H_{0}$ in (\ref{H}) is linear and a simple WKB method suffices. The first energy level is given by the Bohr--Sommerfeld formula:
\begin{eqnarray} 
E_{1} & = & \left(\frac{9m}{8}[\pi\hbar g(1-\frac{1}{4})]^2\right)^{\frac{1}{3}}
\label{WKB_energy_1}\\
& = &\alpha_{1} g^{\frac{2}{3}} \ ; \ n=1, \ 2, \ 3...
\label{WKB_energy_2}
\end{eqnarray}
with $\alpha_{1}=\left(\frac{9m}{8}[\pi\hbar(1-\frac{1}{4})]^2\right)^{\frac{1}{3}}$.
This approximation gives nearly exact value for the first energy level.
\begin{eqnarray}
E_{1}  = 2.23\times 10^{-31} \ \mathrm{(J)}  = 1.392 \ \mathrm{(peV)}
\label{WKB_energy_3}
%\\ &&E_2=2.447 \ {\textrm {peV}}=3.92\times 10^{-31}{\textrm{J}}.\label{ener2}
\end{eqnarray}
as compared to equation (\ref{energy_1}).
Since the perturbation term $H_{1}$ in (\ref{H}) is also linear in $x$ we can combine it with the potential term and rewrite the potential term as
\begin{eqnarray}
V(x) = \tilde{m} g^{\prime} x = \tilde{m} g \left(1 - \frac{\eta \tilde{m}}{\hbar} \right) x 
\label{WKB_energy_4}
\end{eqnarray}
Now using the modified acceleration $g^{\prime}$ from (\ref{WKB_energy_4}) in (\ref{WKB_energy_2}) the approximate shift in the energy value is obtained by first order expansion in $\eta$ as 
\begin{eqnarray}
E_{1} + \Delta E_{1} 
& = & \alpha_{1} g^{\prime}{}^{\frac{2}{3}} 
= \alpha_{1} g^{\frac{2}{3}} \left(1 - \frac{\eta \tilde{m} g}{\hbar}\right)^{\frac{2}{3}} \nonumber\\
& = &  
\alpha_{1} g^{\frac{2}{3}} \left(1 - \frac{2 \eta \tilde{m} g}{3\hbar}\right)\nonumber\\
& = &
E_{1} - \eta \left(\frac{2 E_{1} \tilde{m} g}{3\hbar}\right)
\label{WKB_energy_5}
\end{eqnarray}
Note that in \cite{RB} similar modification of the gravitational acceleration has been made to accommodate the perturbation term in the potential. 
Using the values of $\tilde{m}, g, \hbar\  \mathrm{and}\ E_{1}$ from (\ref{constants}) and (\ref{WKB_energy_3}) we calculate the energy shift $\Delta E_{1}$:
\begin{eqnarray}
\Delta E_{1}  = 2.304\times 10^{-23}\eta \ \mathrm{(J)}  
\label{WKB_energy_6}
\end{eqnarray}
Again this is comparable with (\ref{energy_corrections}). So we get nearly the same upper bound on the time-space NC parameter as in (\ref{eta_bounds_1}) by comparison with the experimental value (\ref{Nesvizhesky_2}):
\begin{eqnarray} 
|\eta| & \lesssim & 2.843\times 10^{-9}\ \mathrm{m^{2}}
%|\eta|&\lesssim& 1.76\times 10^{-61}\ \mathrm{kg^2m^2s^{-2}}\qquad(n=2)~.
\label{eta_bounds_2}
\end{eqnarray}
%____________________________________
\subsubsection{Virial Theorem Method}
%____________________________________
Another simple analytical approach to calculate the energy shift $\Delta E_{1}$ is to use the virial theorem \cite{brau} which implies $\langle T \rangle =\frac{1}{2}\langle V \rangle$ where $T$ and $V$ are kinetic and potential energies, respectively. Hence total energy is given by $ E = \frac{3}{2}\langle V \rangle$. The gravitational potential is $V = \tilde{m} g x$, which gives
\begin{eqnarray}
\langle x\rangle=\frac{2E}{3\tilde{m}g}
\label{virial_1}
\end{eqnarray}  
Now the perturbation term is 
\begin{eqnarray}
H_{1} = \eta\left(\frac{{\tilde{m}}^{2}{g}^{2}}{\hbar}  \right)\langle x \rangle
\label{virial_2}
\end{eqnarray} 
Here using (\ref{virial_1}) we find the energy shift in the first energy level as
\begin{eqnarray}
\Delta E_{1} = - \eta \left(\frac{2 E_{1} \tilde{m} g}{3\hbar}\right)
\label{virial_3}
\end{eqnarray}
which reproduces the same expression for $\Delta E_{1}$ as derived in (\ref{WKB_energy_5}). Hence the upper bound on $\eta$ using the virial theoram method is exactly same as in (\ref{eta_bounds_2}).

This concludes our analysis of the perturbed energy spectrum of the gravitational quantum well problem. This analysis leads to the evaluation of a upper bound on the time space NC parameter in three independent method. 
%%%%%%%%%%%%%%%%%%%%%%%%%%%%%%%%%%%%%%%%%%%%%%%%%%%%%%%%%%%%%%%%%%%%%%%%%%%%
\section{Comparison with existing results}
%%%%%%%%%%%%%%%%%%%%%%%%%%%%%%%%%%%%%%%%%%%%%%%%%%%%%%%%%%%%%%%%%%%%%%%%%%%%
Now that we have a order of magnitude estimation for the time space NC parameter it is instructive to enquire whether it is in conformity with the estimates of other NC parameters reported earlier \cite{bert0, bert1, RB}. In \cite{bert0} the upper bound on the fundamental momentum scale was calculated to be 
\begin{eqnarray}
\Delta p & \lesssim & 4.82 \times 10^{-31}\ \mathrm{kg\ m\ s^{-1}}
\label{bert_scale1}
\end{eqnarray}
Since $E \approx \frac{p_{y}{}^{2}}{2\tilde{m}}$ so 
\begin{eqnarray}
\Delta E \approx \frac{p_{y}}{\tilde{m}}\ \Delta p_{y} = v_{y} \Delta p_{y} \lesssim 31.33 \times 10^{-31} kg\ m^{2}\ s^{-2}
\label{bert_scale2}
\end{eqnarray}
Here we have used the value of $v_{y} = 6.5\ \mathrm{m\ s^{-1}}$ used by the GRANIT experiment group. Using this value of $\Delta E$ in the time energy uncertainty relation $\Delta E \Delta t \geq \hbar$, we find 
\begin{eqnarray}
\Delta t \geq \frac{\hbar }{\Delta E} = 3.38 \times 10^{-4}\ s
\label{bert_scale3}
\end{eqnarray}
Hence uncertainty in time-space sector can be calculated using the results of \cite{bert0} as
\begin{eqnarray}
\Delta x \ \Delta t \sim  3.38 \times 10^{-18}\ m\ s
\label{bert_scale4}
\end{eqnarray}
where following \cite{bert0} we have taken $\Delta x \simeq 10^{-15} \ \mathrm{m}$.
On the other hand in the present paper we have derived the upper bound on the parameter $\eta$ as 
\begin{eqnarray}
\eta = - i \ \left[x^{1}, x^{0}\right] & \lesssim & 2.843\times 10^{-9}\ \mathrm{m^{2}}
\label{bert_scale5}
\end{eqnarray}
Restoring the $c$-factor in (\ref{bert_scale5}) we write the commutator in terms of $x$ and $t$ variables
\begin{eqnarray}
 - i \ \left[x , t \right] = \frac{\eta}{c} = & \lesssim & 9.51 
 \times 10^{-18} \mathrm{m \ s}
\label{bert_scale6}
\end{eqnarray}
Using the generalised uncertainty theorem \cite{jjs} for the commutation relation in (\ref{bert_scale6})
we can write 
\begin{eqnarray}
 \Delta x \ \Delta t \geq \frac{1}{2}\frac{\eta}{c} \sim 4.75 \times 10^{-18} \mathrm{m \ s}
\label{bert_scale7}
\end{eqnarray}
Interestingly the value of the upper bound on the time-space NC parameter as derived here turned out to be consistent with the results of \cite{bert0, bert1, RB}. However, one should keep in mind that this value is only in the sense of an upper bound and not the value of the parameter itself. 

%@@@@@@@@@@@@@@@@@@@@@@@@@@@@@@@@@@@@@@@@@@@@@@@@@@@@@@@@@@@@@@@@@@@@@@@@@@@@@

\section{Conclusions}

%@@@@@@@@@@@@@@@@@@@@@@@@@@@@@@@@@@@@@@@@@@@@@@@@@@@@@@@@@@@@@@@@@@@@@@@@@@@@@
In this paper we have obtained an effective noncommutative quantum mechanics (NCQM) for the gravitational well problem starting from a noncommutative (NC) Schr\"{o}dinger action coupled to external gravitational field. The effective commutative field theory is shown to take the usual form once the proper canonical pair of field variables are identified by a Hamiltonian analysis. The effect of noncommutativity on the mass parameter appears naturally in the process. We reinterprete this one particle field theory as a first quantized theory and obtain an effective noncommutative quantum mechanics (NCQM) for a particle trapped in earth's gravitational field. Interestingly, we observe that the external gravitational field has to be static and uniform in order to get a canonical form of Schr\"{o}dinger equation upto $\eta$-corrected terms so that a natural probabilistic interpretation emerges. 

 The main object of our analysis is to study the gravitational quantum well problem reducing it from a field theoretic setting so that time-space noncommutativity may be included in a natural way.  The singularly important outcome of our calculation is that it is the underlying time-space sector of the NC algebra that is instrumental in introducing non-trivial NC effects in the energy spectrum of the system to first order perturbative level. Following \cite{bert0, bert1} we demand the calculated perturbation in the energy level should be less than or equal to the maximum energy shift allowed by the GRANIT experiment performed at Grenoble \cite{nes1, nes2, nes3}. This comparison leads to an upper bound on the time-space NC parameter. This upper bound is shown to be consistent with the existing upper bound for the fundamental momentum scale in the literature.

%---------------------------------------------------------------------------------------------
\section*{Acknowledgment}

%------------------------
The author would like to acknowledge the hospitality of IUCAA where part of this work has been done. He is also thankful to P.Mukherjee for going through the manuscript and making importan suggestions. Discussions with R.~Banerjee, S.~Samanta and A.~Rahaman are acknowledged.  The author would also like to thank the Council for Scientific and Industrial Research (CSIR), Govt. of India, for financial support.

\end{document}